\documentclass{aa}

\newcommand{\mygal}{NGC\,6822}

\newcommand{\gratis}{{GRATIS}}
\newcommand{\isis}{{ISIS\,2.1}}
\newcommand{\xccdred}{{XCCDRED}}
\newcommand{\allframe}{{ALLFRAME}}
\newcommand{\daophot}{{DAOPHOT}}

\usepackage{graphicx}
\usepackage{txfonts}
\usepackage{color} 

\begin{document}

\title{Variable stars in the dwarf irregular galaxy \mygal\ : the
photometric catalogue\thanks{Based on data collected at the European
Southern Observatory, proposal number 67.B-0557}}

\author{
     L. Baldacci\inst{1,2}
\and L. Rizzi\inst{3,4}
\and G. Clementini\inst{1}
\and E. V. Held\inst{3} 
}

\institute{
INAF - Osservatorio Astronomico di Bologna, via Ranzani 1, I-40127 Bologna, 
Italy \\
\email{lara.baldacci,gisella.clementini@bo.astro.it}
\and
Dipartimento di Astronomia, Universit\`a di Bologna, via Ranzani 1, I-40127 
Bologna, Italy
\and
INAF - Osservatorio Astronomico di Padova, vicolo dell'Osservatorio 5, I-35122 
Padova, Italy\\
\email{held@pd.astro.it}
\and
Institute for Astronomy,
University of Hawaii, 2680 Woodlawn Drive, Honolulu, HI 96822, USA \\
\email{rizzi@ifa.hawaii.edu} (present address) 
}

\offprints{L.~Baldacci}

\date{Received 17 July 2004/ Accepted 19 October 2004}

\abstract{
Deep $B$,$V$ time-series photometry obtained with the ESO Very Large Telescope
has been used to identify variable stars in the dwarf irregular galaxy
\mygal.
We surveyed a 6.8$\times$6.8 arcmin area 
of the galaxy and detected a total number of 390 candidate variables
with the optimal image subtraction technique (Alard \cite{alard00}).
Light curves on a magnitude scale were obtained for 262 of these
variables.  Differential flux light curves are available for the
remaining sample.  In this paper we present the photometric catalogue
of calibrated light curves and time-series data, along with
coordinates and classification of the candidate variables.  A
detailed description is provided of the procedures used to identify
the variable stars and calibrate their differential flux light curves
on a magnitude scale.
}
 
\titlerunning{Variable stars in \mygal\ : the photometric catalogue}
\authorrunning{Baldacci et al.}

\maketitle

\section{Introduction} \label{sez:introduction}
The potential of variable stars in setting the astronomical
distance scale and addressing the origin of stellar populations in
galaxies is demonstrated by the increasing number of surveys devoted
to this subject in several Local Group galaxies.
In a previous paper (Clementini et al. \cite{gis03}) we 
reported the first discovery of RR Lyrae and
other short-period variable stars in the dwarf irregular galaxy 
(dIrr) \mygal, based on deep time-series
photometry obtained with the ESO Very Large Telescope (VLT).
We derived a distance of $(m-M)_{0}$=$23.36\pm$0.17 from the average 
luminosity of the RR Lyrae stars in \mygal\ (Clementini et al. 
\cite{gis03}). 
This paper presents our reduction methods and provides coordinates and
photometry of the candidate variable stars as well as the catalogue of
available light curves.  
This study provides the first comprehensive account of the variable
star content in \mygal, reaching the Horizontal Branch (HB) of the old
population. Previous studies (Hubble \cite{Hu25}, Kayser \cite{ka67},
McAlary et al. \cite{McA83}, Gallart et al. 
\cite{GAV96}, Antonello et al.  \cite{a02}, Pietrzy\'nski et al. 
\cite{Pie04}) only addressed brighter stars (Classical
Cepheids, $V <$ 22 mag), a few of which have been recovered in the
present study.  The properties of the variable stars and their
implications for the stellar populations and star formation history of
\mygal\ will be analysed and discussed in a forthcoming paper (Held et
al. \cite{ev04}).

In Sect.~\ref{sez:observations} and Sect.~\ref{sez:reduction} we describe the 
acquisition, reduction, and absolute photometric calibration of the data.
The procedures employed to detect variable star candidates with the 
optimal image subtraction method (\isis, Alard \cite{alard00}) and to
calibrate their differential flux light curves to a magnitude scale are 
described respectively in Sect.~\ref{sez:identification} and 
Sect.~\ref{sez:calibration}. In Sect.~\ref{sez:variables} 
we discuss the period 
search technique and the classification of
the variables in types, and provide the photometric time-series data. 
In Sect.~\ref{sez:types} we give a general
description of the variable star 
characteristics and discuss their position on the
Color Magnitude Diagram (CMD) of \mygal.  
The atlas of light curves for the  
different types of variable stars with good 
sampling of the light variation is presented in the Appendix. 

\section{Observations} \label{sez:observations}

\begin{figure}[ht]
\resizebox{\hsize}{!}{\includegraphics[bb=66 1 471 388,draft=true]{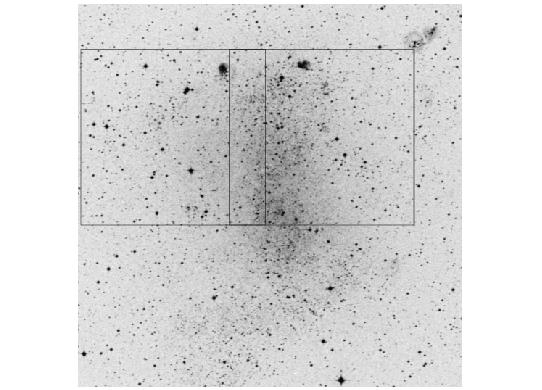}}
\caption{A $15 \times 15$ arcmin Digitized Sky Survey image of 
\mygal\ with our two fields marked by boxes (Field A on the left 
and Field B on the right). North is up and East to the left.}
\label{fig:fig1}
\end{figure}

Observations of \mygal\ were obtained in 3 half nights on Aug.~15, 16,
and 20, 2001 using the focal reducer FORS1 on the ESO VLT/UT3 
(Melipal) telescope at Cerro Paranal, Chile.  All three nights
were photometric and with good seeing conditions.  The standard
resolution collimator was used, yielding a $6.8 \times 6.8$ arcmin
field-of-view with a 2048$\times$ 2048 pixel Textronix CCD having
pixel scale 0.20 arcsec/pixel, read in 4-port mode without
binning. Observations {were obtained} 
in the Johnson-Bessell $B$,$V$, and $I$ filters (ESO Nos. 34, 35
and 37).
We observed two fields covering a
northern portion of the galaxy. Centers of the observed fields are 
$\alpha=19^h 45^m 13\fs3$, $\delta=-14^\circ 45^\prime
55^{\prime\prime}$ (field A) and $\alpha=19^h 44^m 48\fs1$,
$\delta=-14^\circ 45^\prime 57^{\prime\prime}$ 
(field B), respectively.
Field~A is offset from regions of active star formation
in \mygal, while Field~B is located in a central region rich in young
stars.
Figure \ref{fig:fig1} 
shows a $15 \times 15$ arcmin Digitized Sky Survey image of 
\mygal\ with our two pointings outlined by rectangles.
Standard star fields from Landolt (\cite{Landolt92}) were also 
observed in the three nights for photometric calibration.

In this paper we present the results for Field~A, the
only one for which $B$,$V$ times-series data were obtained. 
These consist of 11$B$ and 36$V$
frames. Since the primary goal of our survey was the detection of the
old stellar population in \mygal, if any, scheduling was devised so as
to optimize the detection of variable stars of RR Lyrae type, and
provide a good coverage of the light variation for their typical
periodicities (i.e. periods between 0.2 and 0.8 days, with optimal
efficiency for periods around 0.6 days). Exposure times of 900 s were
adopted in both photometric bands, which represent the best compromise
between signal-to-noise ratio and time resolution for RR Lyrae-like
light curves.  Table
\ref{tab:osservazioni} provides the journal of observations along with
a brief description of the observing conditions during the run.

\begin{table*}[ht]
\begin{center}
\caption{Journal of the photometric observations}
\begin{tabular}{c c c c c c c c c c c}
\hline
\multicolumn{1}{c}{Observing date (UT)} & \multicolumn{1}{c}{HJD-2452100}&
\multicolumn{7}{c}{N.of Observations}&\multicolumn{1}{c}{Photom. cond.}&\multicolumn{1}{c}{Seeing}\\
&&\multicolumn{3}{c}{Field A}&&\multicolumn{3}{c}{Field B}&\\
              &               &$B$   & $V$ &  $I$  && $B$ & $V$  & $I$& & arcsec \\
\hline
Aug.~ 15, 2001& 37.494-37.673 &  4  &  11 &   1  && -- &   1 &  --  & photometric & 0.5-0.8 \\
Aug.~ 16, 2001& 38.490-38.668 &  4  &  12 &  --  &&  1 &  -- &   1  & photometric & 0.7-0.8 \\
Aug.~ 20, 2001& 42.496-42.680 &  3  &  13 &  --  && -- &  -- &   1  & photometric & 0.6-1.0 \\
              &               &     &     &      &&    &     &      &       & \\
\multicolumn{2}{c}{Total}           & 11  &  36 &   1  &&  1 &   1 &   2  &             &       \\
\hline
\end{tabular}
\label{tab:osservazioni}
\end{center}
\end{table*}

\section{Reduction and photometric calibration} \label{sez:reduction}

The FORS1 images were reduced in a standard way using the package
\xccdred\ in IRAF\footnote{IRAF is distributed by the National Optical
Astronomical Observatories, which are operated by the Association of
Universities for Research in Astronomy, Inc., under cooperative
agreement with the National Science Foundation}.  
The photometric reduction was carried out using the \daophot\ and
\allframe\ packages (Stetson \cite{Stetson94}). 
A master
object list was obtained from a stacked image by combining all the
frames regardless of the filter. This was given as input to \allframe\
for simultaneous PSF-fitting photometry of all the individual images.


Although our observing nights were photometric, an independent direct
calibration could not be obtained because our standard stars were
mostly saturated even with the shortest exposure times allowed by the
instrument software.  We therefore used the color terms and zero
points provided by the FORS Web page to convert the instrumental
$b$,$v$, and $i$ magnitudes to the standard Johnson-Bessell
system. The zero points were then refined using secondary standard
stars in the target fields established with Wide Field Imager
observations at the 2.2 m ESO/MPI telescope obtained on photometric
nights and accurately calibrated onto the Landolt (\cite{Landolt92})
system.

The adopted calibration relations are: 

\begin{equation}
V = v + 0.034 \times (B-V) + 27.506
\label{eq:V}
\end{equation}
\begin{equation}
B = b - 0.09 \times (B-V) + 27.207
\label{eq:B}
\end{equation}
The estimated uncertainty of
the zero point calibration is 0.04 mag in both $B$ and $V$.

\section{Identification of the variable stars} \label{sez:identification}
Candidate variable stars were identified using 
the optimal image subtraction technique within the package 
\isis\ (Alard \& Lupton \cite{alardlupton98}, Alard 
\cite{alard00}).
The package was run 
independently on the $V$ and $B$ time-series data.
Images in each photometric band were 
aligned and re-mapped onto the same grid. 
$V$ and $B$ reference images were then constructed
by stacking the 8 $V$ and the 4 $B$ frames with the best seeing.  
Each individual frame of the time series was subtracted from the 
corresponding reference image after convolution
with a suitable kernel to match seeing variations 
and geometrical distortions on the individual images. The last step
was photometry of
the variable objects on the difference images, in terms of
differential fluxes.   
%
%
The interested reader is referred to \isis\ tutorial (available at 
http://www2.iap.fr/users/alard/package.html)
for a description of the relevant parameters and their meaning.

We excluded one of the $V$ frames from the \isis\ master list 
since it is affected by a bright ring produced by light of the guide star
reflected by the telescope optics, which caused serious problems 
to \isis\ background fitting procedure. In 
conclusion, we have 11 and 35 points on 
the $V$ and $B$ light curves, respectively.

The output of \isis\ 
is a median image of all difference frames (known as ``var.fits''), in
which non-variable objects disappear and candidate variable stars
stand out as bright peaks.  Selection of the candidate variables on
the ``var.fits" frames was done by eye, excluding false detections due
to CCD bad columns and hot pixels, satellite trails, and
saturated objects.  \isis\ automatically masks pixels with flux higher
than a threshold defined by the parameter ``saturation''. 
However, since \isis\ works on single pixels, 
only the very central pixels of the saturated stars
are actually masked, while pixels around the saturated stellar core
still appear on the ``var.fits'' image as a large number of false
detections.
%
Since defects of each image of the time series are reflected in the
``var.fits'' image, and because of the high percentage of saturated
objects in our field, only about 10\% of the detections turned out to
be "bona fide" candidate variables. 
After cleaning the preliminary output list, a search performed
on the $V$ ``var.fits'' yielded a master list of 397 candidate
variable stars.
The same search on the $B$ ``var.fits'' yielded a master list of 
286 objects, of which 236 are in common with the $V$ master list.
The smaller number of candidate variables identified in the $B$ ``var.fits''
frame is probably due to (i) the fewer number of $B$ frames, (ii) the intrinsic
faintness of the missing objects in the $B$ band, and (iii) the large  
reddening of \mygal\ which affects the $B$ band more than the $V$ one. 
As a matter of fact, many of the $V$ candidate variables that 
were not counter-identified in the  
$B$ ``var.fits'' master list, have \daophot/\allframe\ measurements
and turned out to be red stars at the tip of the red giant branch
(RGB). 
There are also 50 candidate variables 
in the $B$ \isis\ master list that do not have counterpart in $V$.
Since the data-points in $B$ are insufficient by themselves to derive
reliable light curves, these objects were not further analysed.
The differential flux light curves were produced for each candidate
variable star by running the \isis\ photometry task 
on the list of ``bona fide''
candidate variable stars. This provided differential flux $V$ light
curves for 397 stars and $B$ light curves for 236 of them.

\section{Calibration of light curves} 
\label{sez:calibration}

As underlined in the previous section, \isis\ produces light curves on
a differential flux scale. Yet, light curves on a magnitude scale,
such as obtained by \daophot/\allframe, are needed
to derive the amplitude of light variation, a defining parameter
of pulsation.

On the other hand, the Image Subtraction Method implemented by \isis\
provides superior performances for detection of variable objects in
crowded fields, and the differential flux measurements have usually
lower internal error than PSF-fitting photometry.  Combining the
information coming from both of these powerful tools is of foremost
importance.

Therefore, the candidate variable stars detected by \isis\ were
counter-identified in the \allframe\ $B$,$V$ master catalogue to
transform the differential flux light curves onto a magnitude scale
and place the variables on the CMD of
\mygal.
The \isis\ coordinate system was matched to that of \daophot/\allframe\ 
to produce catalogues with unique identifying numbers and
coordinates. At the end of the whole procedure we were able to match 
302 of the objects flagged by \isis\ as variable sources in the $V$
``var.fits'' frame, 204 of which have counterpart also on the $B$
``var.fits'' frame.
%

The following step was to transform differential fluxes into
(instrumental) magnitudes.  According to the \isis\ tutorial, the
differential fluxes can be transformed using the equation:
\begin{equation}
m_{\star,i}=-2.5\log (F_{\star, {\rm ref}} - \Delta F_{\star,i} )+ C_0.
\label{eq:base}
\end{equation}
where 
the index $i$ runs on the time-series sequence; $m_{\star,i}$ is the 
instrumental magnitude of the candidate variable star in each frame of the 
time-series;
$\Delta F_{\star,i}$ are the differential fluxes produced by \isis;  
$F_{\star,{\rm ref}}$ is the flux of the variable star in the reference image; and 
$C_0$ is a constant which depends on the photometric reduction 
package (in the case of \daophot/\allframe\ $C=25$).
$F_{\star,{\rm ref}}$ can be obtained from the following equation: 
\begin{equation}
F_{\star,{\rm ref}} = 10^{\left(\frac{C_0-m_{\star,{\rm ref}}}{2.5}\right)}
\label{eq:fluxref}
\end{equation} 
where $m_{\star,{\rm ref}}$ is the instrumental magnitude of the star in the 
reference image scaled to the aperture used in \isis\  (the parameter 
``rad\_aper'' in the configuration file 
``phot.data''). 


However, an application of Eqs.~\ref{eq:base} and \ref{eq:fluxref} 
to our data was
not able to correctly reproduce the average magnitude and the
amplitude of the \daophot/\allframe\ light curves for our candidate
variables.
%
%
Similar calibration inconsistencies were encountered by other
users of the Image Subtraction Method (either 
\isis\ or revised versions of this package, as in Pigulski et al. 
\cite{pigulski03}).  


Our procedure allows us to transform the differential
fluxes into instrumental magnitudes by modelling each differential
flux light curve onto that (in magnitudes) produced
with fully calibrated \allframe\ measurements. 
To properly reproduce both the average magnitude of the variable stars
and the amplitude of their light curves,  Eq.~\ref{eq:base} 
has to be  
modified by adding two scaling parameters, $A$ and $C^*$:
\begin{equation}
m_{\star,i}=-2.5\log [F_{\star,{\rm ref}} - A \times \Delta F_{\star,i}]+ (C_0+C^*)
\label{eq:mia}
\end{equation}
Note that the scale parameter $A$ affects the mean luminosity as
well as the amplitude of the light curve.  For $A=1$ and $C^*=0$ we
re-obtain Eq.~\ref{eq:base}.  Hereinafter $C=C_0+C^*$. $A$ and $C$ are
expected to have values close to 1 and 25, respectively. In our
procedure we then make them vary in the range from $0.5$ to $2$ and
from $24$ to $26$ by steps of 0.01.  We then apply Eq.~\ref{eq:mia} to
each data point on the \isis\ differential flux light curve and
look for the $A$ and $C$ pair that, within
the allowed range of values, minimizes the sum:
\begin{equation}
\sum_{i=1}^{i=k}\frac{(m_{i,{\rm ALLF}}-m_{i,{\rm ISIS2.1}})^2}{n}
\label{eq:chi}
\end{equation}
where the $i$ index runs on the time-series sequence (from 1 to 35 in $V$ 
and from 1 to 11 in $B$); 
$m_{i,{\rm ALLF}}$ are the \allframe\ 
magnitudes of each variable star\footnote{Note that to account
for variations in the photometric conditions and airmass
of observation the \allframe\ light curves of the variable stars are 
always differential with respect to a constant comparison star. The
$m_{i,ALLFRAME}$ value in Eq.~\ref{eq:chi} are obtained from the
differential values  
by adding the average \allframe\ 
magnitude of the comparison star 
over the full time-series.}; 
$m_{i,ISIS2.1}$ are the magnitudes 
calculated from the differential fluxes using Eq.~\ref{eq:mia};
and $n$ is the number of acceptable points on the light curve (see below).

The possible presence of spurious data points in
either $B$ or $V$ 
light curves may affect the minimization process. To remove bad
data, an automatic cleaning procedure 
takes into account the photometric error of the \allframe\ measurements and  
performs the rejection of all data points that 
{\it do not} satisfy the condition:
\begin{equation}
(m_{i,{\rm ALLF}}-m_{i,{\rm ISIS2.1}}) < err_{i,{\rm ALLF}} 
\label{eq:errore} 
\end{equation}
where the index $i$ runs on the time-series sequence, $err_{i,{\rm ALLF}}$ is 
the error associated to the \allframe\ 
magnitude measurements $m_{i,{\rm ALLF}}$; 
and 
$m_{i,{\rm ISIS2.1}}$ are the magnitudes 
calculated from the differential fluxes using Eq.~\ref{eq:mia}.  
Since the number of points that satisfy 
Eq.~\ref{eq:errore} varies with the values of the $A$ and $C$ pairs, we only 
considered parameter pairs for which more than the 75\% of light curve 
data points are  
retained. For 
the same reason the sum in Eq.~\ref{eq:chi} 
is also normalized to the number of acceptable points in the light 
curves, $n$.
The procedure was extensively tested on simulated time-series images
of variable stars with characteristics similar to those detected in
\mygal.

\begin{figure*}
\centerline{
\resizebox{7cm}{!}{\includegraphics[bb=23 156 581 700, draft=true]{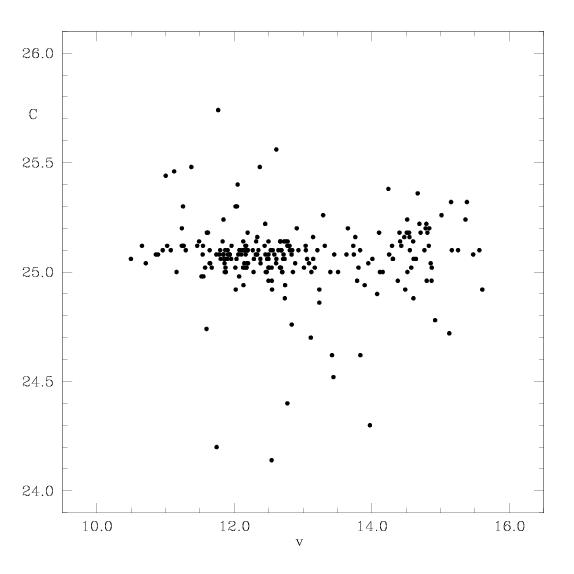}}
\resizebox{7cm}{!}{\includegraphics[bb=23 156 581 700, draft=true]{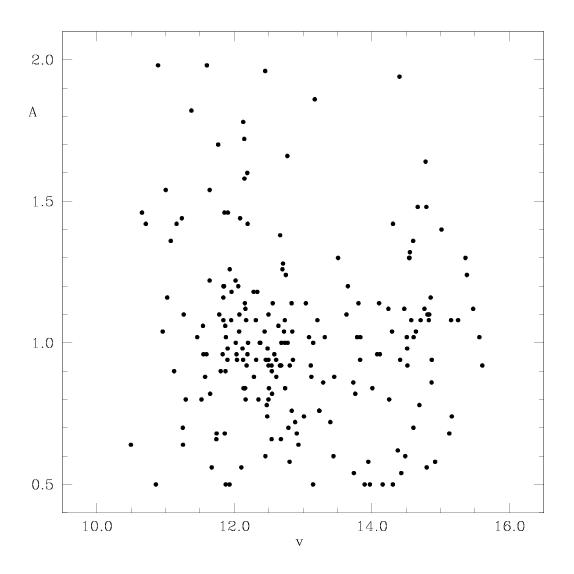}}
}
\centerline{
\resizebox{7cm}{!}{\includegraphics[bb=23 156 581 700, draft=true]{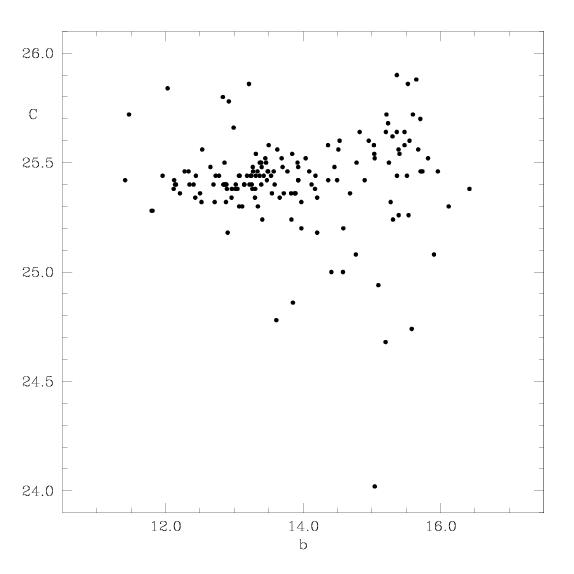}}
\resizebox{7cm}{!}{\includegraphics[bb=23 156 581 700, draft=true]{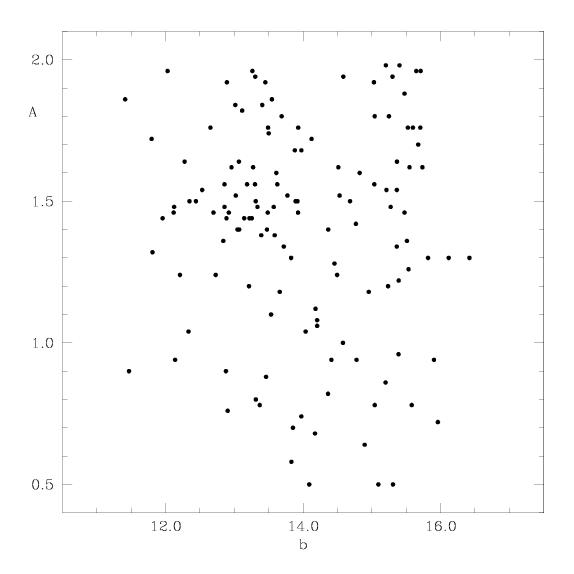}}
}
\caption{Run of $A$ and $C$ values as a function of 
$m_{\star,{\rm ref}}$ in the $V$ band (upper panels) and in the $B$ band (lower 
panels), respectively.}
\label{fig:stat}
\end{figure*}

Figure~\ref{fig:stat} shows the run of the $A$ and $C$ values as
a function of $m_{\star,{\rm ref}}$ in the $V$ band (upper panels)
and in the $B$ band (lower panels). The distribution of the parameter
$A$ is quite sparse in both bands, while $C$ tends to be close to a
mean value of about 25 in $V$, the expected value, and generally
higher in $B$.

\begin{figure}
\resizebox{\hsize}{!}{\includegraphics[bb=30 156 574 696, draft=true]{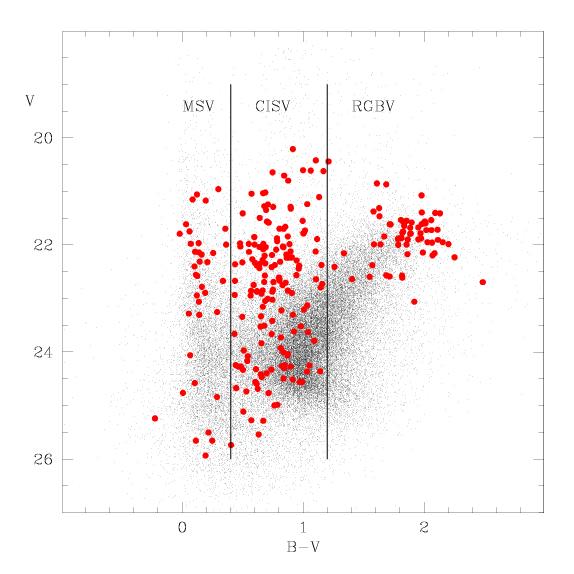}}
\caption{$V$, $B-V$ color-magnitude diagram of Field A in \mygal\, with
the candidate variable stars identified by ISIS2.1 and recovered on the 
\daophot/\allframe\ master catalogue (262 objects) plotted as filled 
circles  (in red in the electronic edition). Vertical lines divide regions corresponding to Main Sequence 
(MSV), Classical Instability Strip (CISV), and Red Giant Branch Variables
(RGBV).}
\label{fig:HR1}
\end{figure}

For candidate variables that were not measured on the reference images
the differential fluxes were calibrated using as a reference the
\allframe\ average magnitudes, if available.
For stars that were not identified  
on the $B$ ``var.fits'' master list produced by \isis, we adopted the \allframe\ 
light curves.
At the end of the calibration process, out of 
the 397  
candidate variable stars originally identified by \isis\, 7  
did not show significant variations and were discarded. 
Therefore the number of ``bona-fide'' variable star candidates 
is 390. 
Light curves on a magnitude scale were obtained in $B$, $V$ for  
262 objects, while only differential flux 
light curves were obtained for further 128 variables. 
The latter include 
candidate variable stars that were neither identified in the 
reference images nor recovered in the \allframe\
master catalogue, along with objects for which we could not find a 
pair of $A$ and $C$ values satisfying Eq.~\ref{eq:errore}  
for the required minimum number of 
points on the
light curve. 

\section{The ``bona fide'' variable stars} \label{sez:variables}

\subsection{The catalogue}

\begin{table*}
\begin{center}
\caption{Variable stars in \mygal}
\label{tab:tabellone}
\begin{tabular}{lrcccclcc}
\hline
IDr & ID~~ & $\alpha$ (2000)& $\delta$ (2000) & $V$ & $\mathbf {B-V}$  & ~Type & Cross-id.&Notes\\
    &    &                &                 &(mag)& (mag)&        &     \\ 
\hline
V1 &  lc282 & 19:45:24.84 & -14:49:13.26  &   $-$  &    $-$  & $-$  &	     & 1\\
V2 &  1020  & 19:45:05.18 & -14:49:13.22  & 22.153 &   2.087 & RGBV &	     & 3\\
V3 &  1140  & 19:45:18.23 & -14:49:12.95  & 21.402 &   2.092 & RGBV &	     & 3\\
V4 &  1453  & 19:45:05.41 & -14:49:09.80  & 22.732 &   0.827 & CISV &	     &  \\
V5 &  lc208 & 19:45:25.38 & -14:49:09.10  &  $-$  &    $-$  & $-$  &	     & 1\\
V6 &  1753  & 19:45:02.95 & -14:49:07.48  & 22.694 &   2.485 & RGBV &	     & 3\\
V7 &  lc267 & 19:45:24.91 & -14:49:06.45  & $-$   &    $-$  & $-$  &	     & 1\\
V8 &  2667  & 19:45:23.61 & -14:49:02.31  & 20.642 &   0.747 & CISV & cep047 & 6\\
V9 &  2572  & 19:45:21.92 & -14:49:02.32  & 21.609 &   1.981 & RGBV &	     & 3\\
V10&  2739  & 19:45:12.23 & -14:49:01.81  & 22.230 &   2.252 & RGBV &	      & 3\\
V11&  lc4682& 19:45:02.74 & -14:48:59.95  & $-$    &    $-$  & $-$  &	      & 1\\
V12&  lc4301& 19:45:04.43 & -14:48:57.99  & $-$    &    $-$  & $-$  &	      & 2\\
V13&  3607  & 19:45:26.17 & -14:48:56.44  & 22.565 &   1.687 & RGBV &	      & 3\\
V14&  3766  & 19:45:19.55 & -14:48:56.51  & 21.547 &   1.855 & RGBV &	      & 3\\
V16&  3518  & 19:45:09.25 & -14:48:56.09  & 24.440 &   0.586 & CISV &	      & 3\\
V17&  4220  & 19:45:12.95 & -14:48:52.91  & 22.378 &   1.572 & RGBV &	      & 3\\
V18&  lc1972& 19:45:14.38 & -14:48:49.43  & $-$    &    $-$  & $-$  &	      & 1\\
V19&  4721  & 19:45:09.33 & -14:48:48.83  & 22.120 &   0.896 & CISV &	      &  \\
V20&  lc1109& 19:45:18.51 & -14:48:47.58  & $-$    &    $-$  & $-$  &	      & 2-7\\
V21&  lc3116& 19:45:09.78 & -14:48:47.52  & $-$    &    $-$  & $-$  &	      & 2\\
V22&  5365  & 19:45:09.89 & -14:48:45.64  & 22.311 &   0.145 & MSV  &	      & 6\\
V23&  lc1959& 19:45:14.43 & -14:48:45.14  & $-$    &    $-$  & $-$  &	      & 2\\
V24&  5659  & 19:45:09.60 & -14:48:44.12  & 21.225 &   0.104 & MSV  &	      &  \\
V25&  5780  & 19:45:15.04 & -14:48:43.72  & 21.730 &   1.015 & CISV &	      & 3\\
V26&  5921  & 19:45:11.73 & -14:48:42.49  & 21.596 &   2.011 & RGBV &	      & 3\\
V27&  6138  & 19:45:00.88 & -14:48:41.44  & 23.219 &   0.934 & CISV &	      &  \\
V28&  6555  & 19:45:06.47 & -14:48:39.12  & 23.968 &   0.762 & CISV &	      &  \\
V29&  6869  & 19:45:03.87 & -14:48:36.42  & 22.743 &   0.865 & CISV &	      &  \\
V30&  lc1909& 19:45:14.73 & -14:48:36.22  & $-$    &    $-$  & $-$  &	      & 1\\
\hline
\end{tabular}
\end{center}

Table~\ref{tab:tabellone} is presented in its entirety in the electronic version of the Journal.
A portion is shown here for guidance regarding its form and content.\\
Notes. \\
1 - Differential flux light curve, only $V$ band available\\
2 - Differential flux light curves, both $B$ and $V$ bands available\\
3 - Not found on \isis\ $B$ ``var.fits" frame, \daophot/\allframe\ $B$ light curve\\
4 - \isis\ light curves are very poor, we used the \daophot/\allframe\ $B,V$ light curves\\
5 - Not found on \isis\ $B$ ``var.fits" frame, \daophot/\allframe\ $B$ light curve.
    Not measured on the $V$ reference image, calibration performed using the \daophot/\allframe\ $V$ average magnitude.\\ 
6 - Not measured on the $B$,$V$ reference images,
    calibration performed using the \daophot/\allframe\ $B,V$ average magnitudes.\\ 
7 - Differential flux light curves calibrated according to the procedure described in Sect.~\ref{sez:fluxes}\\ 
\end{table*}

Coordinates for the {\it ``bona fide"} variable stars 
(390 objects),
are provided in Table~\ref{tab:tabellone}, where in 
Col.~1 we give a running identifier (IDr) defined with the stars
ordered by increasing declination; Col.~2 gives either the
\isis\ {\it ``lc"} number or the \daophot/\allframe\ ID depending on
the available data; Col.~3 and 4 provide $\alpha$ and $\delta$
coordinates (J2000.0); and Cols.~5 and 6 give the average $V$
magnitudes and $(B-V)$ colors for stars with calibrated light curves.
For these stars, 
a coarse classification
mainly based on the color and position on the color-magnitude  
Diagram is provided in Col.~7. We define Main Sequence
Variables (MSV) all objects with $B-V\leq0.4$, Classical Instability Strip 
Variables (CISV) stars with $0.4\leq B-V\leq1.2$, and Red Giant Branch Variables 
(RGBV) objects with colors $B-V\geq1.2$. 
These three groups contain 36, 160, and 66 variable stars, respectively.
%
Finally, Col.~8 provides a cross-identification of the bright variables 
in common with Antonello et al. (\cite{a02}) and 
Pietrzy\'nski et al. (\cite{Pie04}); and 
Col.~9 gives a flag
indicating the type of available light curves, or the
light curve calibration process (no flag indicates
variables with light curves calibrated from \isis\ differential fluxes
according to the general procedure described in
Sect.~\ref{sez:calibration}). 
%
%
Figure~\ref{fig:HR1} shows the location of the 262 variables 
with magnitude-calibrated light curves
on the CMD of \mygal. The variables are
plotted according to their average \daophot/\allframe\ magnitudes and
are divided into the 3 above mentioned classes.
The MSV sample is likely to contain: $\beta$ Cepheids,
$\delta$ Scuti and SX Phoenicis stars and binary
systems (see Sect.~\ref{sez:MSV});  
RR Lyrae, Anomalous Cepheids (ACs), 
Population II Cepheids (W Virginis and BL Herculis stars) and 
Classical Cepheids ($\delta$ Cepheids) would fall in the CISV sample
(see Sect.~\ref{sez:CISV}); and the
Miras, the Semiregular variables and the Small Amplitude Red Giants (SARGs)
in the RGBV sample (see Sect.~\ref{sez:RGBV}).

\subsection{Period search and average properties} \label{sez:periods}

The light curves of the variable stars were studied 
using the program \gratis\ (GRaphical Analyzer of TIme Series; see
Clementini et al. \cite{gis00}, and Di Fabrizio \cite{df99}, for details).
Using this program a first estimate of the period was obtained 
from the 
Lomb periodogram (Lomb \cite{lomb76}, Scargle \cite{scargle82}); 
then we refined the period definition and determined the
best fitting model of the star light variation using 
truncated Fourier series 
(Barning \cite{barning63}).
Given the short time baseline of our observations (3 half nights
spread over a total interval of 5 nights), good sampling of the light
curves is achieved and periods can be reliably determined only for
stars with periodicities shorter than 2-3 days.
Moreover, although the study of period can be performed both for
variables with light curves on magnitude scale and for stars with
differential flux light curves, only for the former a full
characterization of the light variation in terms of amplitude and
average magnitude can be obtained, thus allowing us to locate the star
on the CMD and classify it in type.
%
For these reasons our efforts were mainly devoted to study  the 
short-period variable stars with magnitude-calibrated light curves 
(see Sect.~\ref{sez:mags}) and only for a few objects 
with promising time series differential fluxes 
an analysis of the differential flux light curves was attempted 
(Sect.~\ref{sez:fluxes}). 

\subsection{Variable stars with light curves on a magnitude scale} \label{sez:mags}
Time series photometry on the magnitude scale is available for 262 of the
variable stars detected in \mygal.
The instrumental-magnitude light curves of each variable were
calibrated to the standard $B$,$V$ Johnson system by modelling the
color variation along the star cycle with the \gratis\ best-fitting
Fourier algorithm and then using the calibration Eqs.~\ref{eq:V},
\ref{eq:B}.
%
The sampling of the 
$B$ data points, although sparser than that of the $V$ observations, was
sufficient to constrain the color quite well. Individual 
$B$,$V$ photometric 
measurements are provided in Table \ref{tab:fotometria}: 
for each star we give 
the running IDr, the \daophot/\isis\ identification number, the variable 
type (when a classification was possible), 
the Heliocentric Julian Day (HJD) of the observations and 
the corresponding $B$, $V$ magnitudes. 

\begin{table}[ht]
\begin{center}
\caption{$B$, $V$ photometry of the variable stars with 
light curves in magnitude scale (262 objects) 
\label{tab:fotometria}}

\begin{tabular}{c c  c c}
\hline
\multicolumn{4}{c}{Variable star 27 $-$ID 6138 $-$ Cepheid} \\
HJD         & $B$ &HJD	    & $V$ \\
($-$2452100)  & (mag) &($-$2452100) & (mag)\\
\hline
37.5107 &  23.875 & 37.4995  & 23.060 \\
37.5663 &  23.755 & 37.5218  & 23.019 \\
37.6216 &  23.804 & 37.5441  & 22.994 \\
37.6566 &  23.923 & 37.5552  & 22.936 \\
38.5058 &  24.666 & 37.5774  & 22.899 \\
38.5612 &  24.531 & 37.5884  & 22.857 \\
38.6166 &  24.487 & 37.5995  & 22.853 \\
38.6513 &  24.375 & 37.6105  & 22.851 \\
42.5119 &  24.425 & 37.6345  & 23.038 \\
42.5855 &  24.472 & 37.6455  & 23.044 \\
42.6632 &  24.367 & 37.6678  & 23.043 \\
\hline
\end{tabular}
\end{center}

A portion of Table \ref{tab:fotometria} is shown here for guidance
regarding its form and content. The entire catalogue of time series
data is available upon request from the authors.
%

\end{table}  

Evenly sampled light curves enabling a reliable estimate of the period are
available for  
69 of the variable stars in this sample. As expected from our observing
strategy, they almost entirely
belong to the CISV class and include: 18 RR Lyrae stars
(16 $ab-$, 2 $c-$type), 45 Cepheids and 6 Binaries.
We add to this number 
further 8 RR Lyrae variable stars with differential flux
light curves approximately calibrated onto a magnitude scale
using the procedure outlined in Sect.~\ref{sez:fluxes}.
Best fitting models based on \gratis\ truncated Fourier series 
algorithm were computed for all these variables  with a number of 
harmonic varying from 1 to 5. Mean magnitudes computed as 
intensity-averaged means over the full variation cycle, and 
amplitudes of the light variation (computed as the difference 
between maximum and minimum of the best fitting model)  were
derived from the light curves.
The results are provided in Tables~\ref{tab:rr}, 
\ref{tab:cc}, \ref{tab:bb}, for 
RR Lyrae stars, Cepheids and eclipsing binaries, respectively.
In these tables, Col.~1 is the running identifier (IDr), 
Col.~2 is the \daophot/\allframe\ or the \isis\ ID; 
Col.~3 gives the variable type, with {\it ab} and {\it c} for 
fundamental and first overtone RR Lyrae stars, respectively;
C for Cepheids (including Anomalous, Low Luminosity and 
Classical Cepheids, see Sect.~\ref{sez:CISV}), and EB for the 
eclipsing binary systems. Columns~4 and 5 
provide periods and epochs (corresponding to the HJD of maximum 
light for the pulsating variables, and to the deeper minimum light
for the binaries). Column~6 gives the numbers of $B,V$ 
data-points available on the 
curves; Cols.~7 and 8 are the intensity-averaged $B,V$ mean magnitudes, 
Cols.~9 and 10 the corresponding amplitudes; Cols.~11 and 12 
provide the residuals 
with respect to light curves obtained from \gratis\ 
best fitting models, and finally remarks
on individual objects are given in Col.~13. 
An atlas of light curves for this subsample of 69 variable stars 
is provided in the Appendix. 

\subsection{Variable stars with light curves in differential flux} 
\label{sez:fluxes}

\begin{figure*}
\resizebox{\hsize}{!}{\includegraphics[bb=48 165 503 685, draft=true]{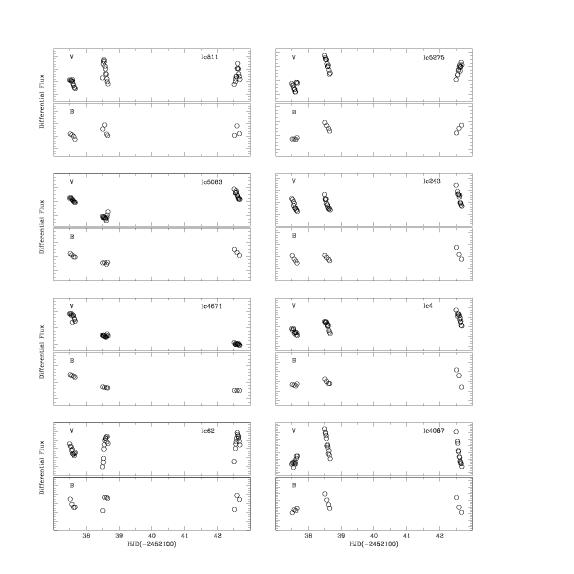}}
\caption{Examples of light curves in differential flux {\it vs.}
HJD of observation for variable stars that we could not calibrate 
onto a magnitude scale
and do not have reliable period determinations.}
\label{fig:c_jd}
\end{figure*}

Among our sample of ``bona fide'' variable stars, 128 objects
only have differential flux light curves, either because they lack
reliable \daophot/\allframe\ time series photometry, or because we
could not measure their magnitudes on the references images.  These objects 
represent about 1/3 of our sample. Some of these variables are located
near bright and/or saturated objects, or are close to the edges of the
frames, hence they could not be reliably measured by
\daophot/\allframe.  However, their vast majority does not have
reliable \daophot/\allframe\ magnitudes because of 
the severe crowding conditions in our field and possibly because they
are very faint, hence close the detection limit of our photometry. 
Examples of light curves (in
differential flux {\it vs.} HJD of observation) for some of these
objects are shown in Fig.~\ref{fig:c_jd}.

\begin{figure*}
\resizebox{\hsize}{!}{\includegraphics[bb=46 540 501 682, draft=true]{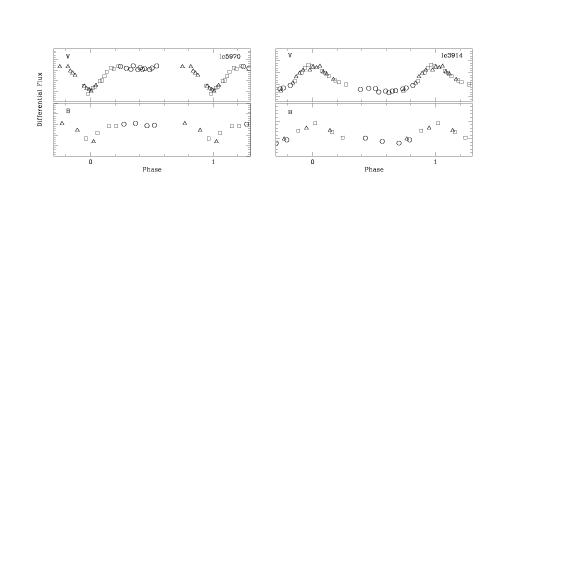}}
\caption{Examples of differential flux light curves for variable stars 
with accurate period determinations but that we could  not calibrate
onto a magnitude scale. 
Different symbols are used for data-points corresponding to the three nights 
of the run. 
Left panel: 
binary system with $P$=0.588 days; right
panel: pulsating variable with $P$=0.405 days.}
\label{fig:c_fase}
\end{figure*}

We ran  \gratis\ on the 
differential flux time series of some of these objects and were
able to determine periods for 10 of them: star lc5970
was found to be a  
binary system with period $P$=0.588 days 
(see Fig.~\ref{fig:c_fase}, left panel). Other 9 objects have 
light curves and periods 
consistent with 
variables of RR Lyrae type (8 $ab-$type RR Lyraes 
and 1 possible $c-$type RR Lyrae star, see right panel of 
Fig.~\ref{fig:c_fase}). 
%

Since the faintest variables measured in \mygal\ with
\daophot/\allframe\ are  RR Lyrae stars, we have attempted a
magnitude calibration of the differential flux light curves of these 9
objects under the assumption that they are indeed RR Lyrae stars.
By assuming as their reference magnitude the mean magnitude of the  
RR Lyrae stars in \mygal,  
and using as template amplitudes those of the \mygal\ RR Lyrae stars
with similar periods, we were able to calibrate the light curves of 8
of these stars with acceptable pairs of $A$ and $C$ values.
Instead, this procedure failed with 
the RRc-like variable, very likely because this is not an RR Lyrae star. 
The light curves obtained for these 8 candidate {\it ab-}type
RR Lyrae stars from this approximate calibration to a magnitude scale  
are shown in Fig.~\ref{fig:appb2}, 
while their 
average quantities are listed in the bottom part of Table \ref{tab:rr}.


\section{Classification of variable stars in \mygal}
\label{sez:types}
\begin{figure}
\resizebox{\hsize}{!}{\includegraphics[bb=40 163 577 697, draft=true]{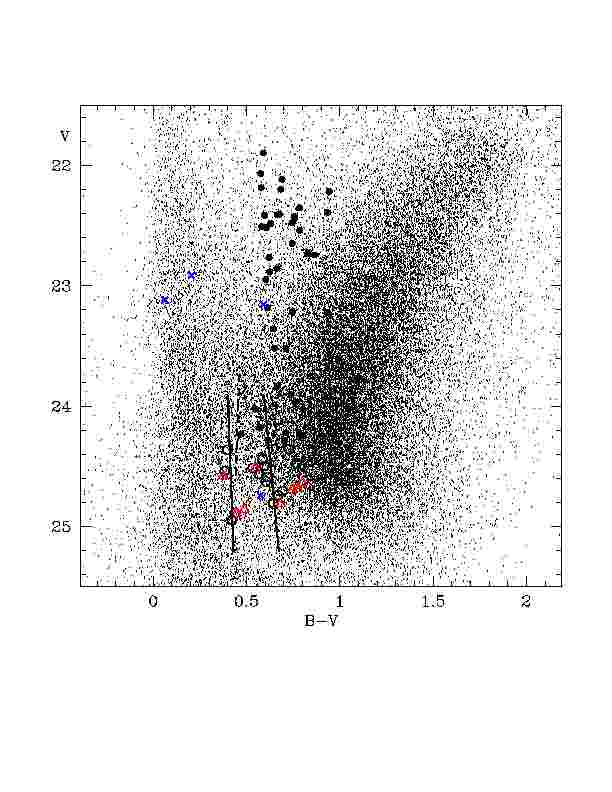}}
\caption{$V$, $B-V$ color-magnitude diagram of Field A in \mygal\ where 
the variable stars with well-sampled light curves are plotted
according to their intensity-averaged magnitudes and colors.  {\it
Filled circles} are Cepheids, {\it open circles} are $ab-$ and $c-$type RR
Lyraes, {\it open triangles} (in red in the electronic version of the
Journal) are the possible $ab-$type RR Lyraes recovered from the
differential flux samples; and {\it crosses} 
(in blue in the electronic version of the
Journal) are binary systems. 
Also shown are the RR Lyrae instability strip for the globular
cluster M\,3 ({\it solid lines}), and the boundaries of the
theoretical pulsational instability strip for 1.5 $M_\odot$ models
({\it short-dashed lines}; see text).  }
\label{fig:HR2}
\end{figure}

Along with the coarse classification in Sect. \ref{sez:variables}
and Fig. \ref{fig:HR1}, as
an aid to classify the confirmed variable stars in \mygal, we
have plotted on the galaxy CMD the location of the subsample of 69
variables with calibrated light curves and good phase coverage
(Fig.~\ref{fig:HR2}). Variable stars are plotted according to their
intensity-averaged mean magnitudes and colors.
For comparison, we also show in Fig.~\ref{fig:HR2} the edges of
 the RR Lyrae instability strip of the globular cluster M~3 (Corwin \&
 Carney \cite{cc01}), shifted to the reddening and distance modulus
 appropriate for \mygal\ (solid lines; 
see also Clementini et al. \cite{gis03}). 
 The short-dashed lines represent the boundaries of the theoretical pulsational
 instability strip for 1.5 $M_\odot$ models from Bono et
 al. (\cite{bo97}), transformed to the observational plane using the
 model atmospheres of Castelli, Gratton, \& Kurucz (\cite{cgk97}).
 As already noted the variables shown in Fig.~\ref{fig:HR2} 
 almost entirely belong to the class of the classical instability
 strip pulsating variables (CISV).

\subsection{The Classical Instability Strip Variables - CISV} \label{sez:CISV} 
The CISV sample includes a total of 160 variables. We have classified 
and determined periods for 65 of them 
(among which we found 2 binary systems),  i.e. for about 
60\% of the variables fainter than $V\sim 23$ mag (hence with shorter period) 
in this group.
There are 18 confirmed plus 8 likely RR~Lyrae stars in this sample.
Basic information and average properties for these stars are summarized in 
Table \ref{tab:rr}, while their light curves are shown in 
Figs.~\ref{fig:appb1} and \ref{fig:appb2} in the Appendix.
Out of these 18 RR~Lyrae stars, 
10 fall within the edges of the instability  strip
of M\,3. Their average magnitude is 
$\langle V({\rm RR}) \rangle = 24.66 \pm 0.17$ mag, 
where the error is the standard deviation of the data.  The average
period of the total sample of 24 {\it ab-}type RR Lyrae stars is
$<P_{ab}>=0.605 \pm 0.036$ days, and $<P_{ab}>=0.611 \pm 0.029$ days
if the 8 additional RRab's are discarded.  These revised values are
consistent with those published in Clementini et al. (\cite{gis03}).

Eight of the RR Lyrae variables lie outside the edges 
of the M3 instability strip, two on the blue and six on the 
red side of the strip (see Fig.~\ref{fig:HR2}). All these stars have been flagged 
in Table \ref{tab:rr}.
Stars \#26506 and 62059 are marginally bluer than 
 the blue edge, the former being also slightly overluminous.
The rather blue color of  \#62059 is probably caused by the poor 
sampling of its light curves.
Out of the six RR Lyraes redder than the red edge of the strip, two are 
also slightly
overluminous in $V$. The $V$ amplitudes of some of them are also too small 
compared to the $B$ amplitudes. A number of these variables 
could be blended with 
red giant stars unresolved in the $V$ band.   
In their study of RR~Lyrae stars in the LMC, Di
Fabrizio et al. (\cite{df04}) discuss in detail the effects of
blending RR~Lyrae variable stars with main sequence and red giant
stars (Sect.~3.2 of that paper). The amplitudes of the light variation
is reduced, and the variable star becomes redder/bluer depending on
the color of the blending star, and overluminous by an amount
depending again on the magnitude of the blending star.
Crowding in our \mygal\ field is very high, hence blending is a very
plausible cause of the too red/blue colors of the RR Lyrae stars
falling outside the instability strip.
A further explanation of the red color of RR~Lyrae stars is the
existence of internal differential reddening in \mygal. For instance,
the two faintest RR Lyrae stars lying slightly outside the strip
(stars \#42860 and \#13845) can be easily brought inside the strip
by assuming a moderate differential reddening, 0.02-0.03 mag.  This
amount of differential reddening is perfectly consistent with the
variations in the internal extinction
inferred 
from the 60$\mu$m emission measured in \mygal\  by the IRAS satellite (Rice \cite{ri93}).
%


 


The CMD in Fig.~\ref{fig:HR2} shows an almost continuous distribution of
variable stars with periods in the range from $\sim 0.36$ to $\sim 2$ 
days, filling the region of the diagram above the HB, beginning just
about 0.3 mag up to 2.7 mag brighter than the average luminosity level
of the RR Lyrae stars, and with increasingly redder mean colors.
The basic properties of these stars are summarized in 
Table \ref{tab:cc}, while 
their light curves are shown in Fig.~\ref{fig:appb3}.
Clementini et al. (\cite{gis03}) indicated the 20 variables with mean
magnitude in the range 23.0$< V \la $24.3 as 
Low Luminosity Cepheids (LL-Cepheids). These stars, flagged as LL-C in 
Table \ref{tab:cc},
are characterized by small
amplitudes (0.1$< A_V \la$ 0.8 mag), short periods overlapping with the 
period distribution of the
RR Lyrae stars, 
and luminosities from only a few tenths to about 1.6 mag brighter than
the average magnitude of the RR Lyraes. The properties of the
LL-Cepheids are discussed more in detail in Held et
al. (\cite{ev04}) -- we only note here that their characteristics are
consistent with the theoretical predictions for Anomalous Cepheids by
Bono et al.  (\cite{bo97}) and with the more recent theoretical
pulsation models by Marconi, Fiorentino, \& Caputo (\cite{mfc04}).

Brighter than the LL-Cepheids ($V <$ 23 mag) we find the vast majority
of longer period Cepheids (25 stars). The average magnitudes and
colors of these variables are in good agreement with the theoretical
strip found for the SMC short-period Cepheids with $Z=0.004$ and
masses in the range $3.250 \leq M/M_{\sun} \leq 7.0$ (M. Marconi 2004,
private communication), once shifted to the average reddening and
distance modulus adopted for \mygal.

Among the LL Cepheids there are 3 objects (labelled with a double ``r"
in Table \ref{tab:cc}) with very red colors ($B-V \sim 0.9 - 1.1$
mag), that fall largely outside the theoretical instability strip for
Anomalous Cepheids. They all lie either in very crowded areas near the
centre of \mygal\ or close to active star formation zones. Other 11
variables among the LL and Classical Cepheids are found slightly
outside the red edges of their respective instability strips (they are
flagged with a single ``r" in Table \ref{tab:cc}).
As for the RR Lyrae stars, several of them lie in regions of \mygal\
probably 
affected by differential reddening, in a few other cases they could be
unresolved blends, as indicated by the high value of the shape
parameter {\it SHARP} in \daophot/\allframe/ photometry. 


Finally, we note that two of our brighter candidate variables with
only partial coverage of the light curves, 
namely stars  \#16070 and 22796,
correspond to the Classical Cepheids V3186 and V3736 
in Antonello et al. (\cite{a02}) and to cep061 and cep025 in 
Pietrzy\'nski et al. (\cite{Pie04}), for which both
authors find periods of  $\sim 4.6$ and $\sim 8.9$ days,
respectively. 
We have further 11 bright variables in common with  
Pietrzy\'nski et al. (\cite{Pie04}) all classified Classical
Cepheids by these authors, 9 classified Classical Cepheids, one MSV (V78) and 
one RGBV 
(V150) for us.
The cross-identification of V150 with cep052 in Pietrzy\'nski et al. list
is doubtful; also uncertain is that of their
cep036 (see column 8 of 
Table\ref{tab:tabellone}).


\subsection{The Main Sequence Variables - MSV} \label{sez:MSV}
The MSV sample contains 36 variable stars. 
In this region of the CMD we expect to find binary systems, 
$\beta$ Cepheids, Be stars, slowly-pulsating B variables (SPB), 
$\delta$ Scuti and SX Phoenicis variables
(see Sterken \& Jaschek \cite{sj96}  for a description of the
characteristics of these different types of variables). 
The  $\delta$ Scuti and SX Phoenicis stars are 
generally found below the Horizontal Branch from luminositites
$M_V \sim$ 3.0 mag to
$M_V \sim$ 0.5 mag (Breger \cite{bre00}),
corresponding to $V \ga$ 24.5 mag in \mygal. Hence, they  
are generally below our detection limit. Indeed, we only have a few 
MS candidate variables fainter than $V \sim $ 24 mag, 
some of which have large 
amplitudes,
 but also have very large errors.

$\beta$ Cepheids, Be stars and SPB variables share approximately the 
same area on the main sequence, 
at luminositites from 
$M_V \sim -$1.5 mag to
$M_V \sim -$4.5 mag (Pigulski \& Kolaczkowski \cite{pk02}), roughly 
corresponding to $V< 23$ mag in \mygal. This is the region 
where many of our MS variables have been found.
No such variable stars were known 
in external galaxies until the recent discovery of many candidate  
$\beta$ Cepheids in the Large Magellanic Cloud (LMC, Pigulski \& Kolaczkowski 
\cite{pk02}, Kolaczkowski \& Pigulski \cite{kp04}). Their typical periods 
are $P <$ 0.3 days. Their amplitudes are 
very small: $A_V <$ 0.1 mag, and often of the order of a few hundredths of magnitudes
(Pigulski \& Kolaczkowski \cite{pk02}). 
The Be stars have longer
periods in the range $0.4 < P < 3$ days and amplitudes $0.01 < A_V < 0.3$ mag.
Finally, the SPB variables have $P \sim 1 - 4$ days, and $A_V \sim 0.01$ mag.
Among these types of variables, only those with longer period and/or larger amplitude 
can be detected at the distance of \mygal.
However, if they are present in our MSV sample, 
the short time interval covered by our observations 
makes their identification very difficult.    
Indeed, the vast majority of our MS variables 
have light curves recalling those of eclipsing binary systems. However, the
phase coverage is too small and our period definition for these stars is 
strongly limited by alias problems.
Light curves for the few (4) MS eclipsing binaries for which we were
able to determine reliable periods
are shown in Fig.~\ref{fig:appb4}.
%
%

\subsection{The Red Giant Branch Variables - RGBV}  \label{sez:RGBV}

\isis\ detected 66 candidate
variables lying around the tip of the RGB of \mygal\ (see 
Fig.~\ref{fig:HR1}). These stars are likely Miras, Semiregular variables, and
Small Amplitude Red Giants (SARGs) as those recently discovered in 
large numbers in the
Galactic Bulge (Wray, Eyer \& Paczy\'nski \cite{wep04}) and in the
Magellanic Clouds (Wood et al. \cite{w99}, Soszy\'nski et al. 
\cite{sosz04}).
%

\begin{figure*}
\resizebox{\hsize}{!}{\includegraphics[bb=46 540 501 682, draft=true]{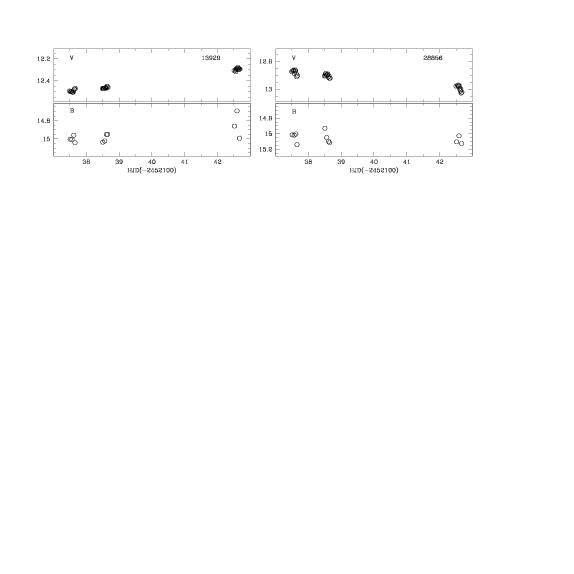}}
\caption{Examples of light curves in 
instrumental magnitude {\it vs.} HJD of observations for
variable stars at the tip of the RGB. 
}
\label{fig:c_rgb}
\end{figure*}

We were not able to determine periodicities for any 
of these variables, since the time interval spanned by our observations is 
too short to monitor their long term variations. 
%
In fact, expected periods are $P \sim 50$, $P \sim 100-200$, and 
$10 < P < 100$ days, for the Miras, the Semiregular and the SARG variables,
respectively.  However, thanks to the capability of the image
subtraction technique we were able to detect the
irregular, small amplitude ($A_V \leq 0.10-0.15$ mag), short time
scale variations of these objects, in particular of some SARG variables.
%
Some examples of time series data for the RGB variables are shown 
in Fig.~\ref{fig:c_rgb}.

\begin{table*}
\begin{center}
\caption{Mean quantities for RR Lyrae stars.}
\label{tab:rr}
\begin{tabular}{lrrc cc@{\hspace{20pt}}l @{\hspace{10pt}}lrr ccc}
\hline
IDr &ID    & ~~~Type &   $P$    & Epoch     &  $N$   &$<B_{int}>$ &$<V_{int}>$&
$A_B$~~~&$A_V$~~~& $resB$  & $resV$ & Notes \\
&      &         & (days) &($-$2452000)& ($B$,$V$) & (mag)           & (mag)          &  (mag)    &  (mag)   & (mag) & (mag) & \\
\hline
V141 & 25119  & c   & 0.369  &  137.6455  & 11,34  &  25.245  &  24.476  &   0.639  &   0.399  &  0.098 &  0.065  &  r\\
V228 & 40170  & c   & 0.406  &  137.6000  &  9,31  &  25.228  &  24.625  &   0.364  &	0.228  &   $-$  &  0.037  &  \\
V367 & 63229  & ab  & 0.575  &  136.9956  & 10,30  &  25.398  &  24.705  &   1.228  &	0.875  &   $-$  &  0.064  &  \\
V283 & 49388  & ab  & 0.577  &  136.4454  &  8,34  &  25.367  &  24.945  &   1.509  &	1.152  &   $-$  &  0.098  &  \\
V241 & 43139  & ab  & 0.585  &  136.8274  & 11,33  &  25.082  &  24.355  &   1.013  &	0.529  &  0.033 &  0.034  &  r,o\\
V152 & 26506  & ab  & 0.591  &  136.6950  & 10,35  &  24.759  &  24.362  &   1.213  &	0.771  &   $-$  &  0.033  &  b,o\\
V77  & 14560  & ab  & 0.594  &  136.5229  & 10,35  &  25.327  &  24.889  &   0.998  &	0.776  &   $-$  &  0.058  &  \\
V210 & 35720  & ab  & 0.601  &  136.6751  & 11,35  &  25.046  &  24.512  &   1.012  &	0.667  &  0.042 &  0.036  &  \\ 	 
V200 & 34017  & ab  & 0.602  &  136.5471  &  8,32  &  25.452  &  24.809  &   1.477  &	0.860  &   $-$  &  0.054  &  \\
V394 & 67718  & ab  & 0.603  &  136.3825  & 11,30  &  25.290  &  24.462  &   0.889  &	0.662  &  0.022 &  0.044  &  r\\
V332 & 57462  & ab  & 0.605  &  136.0641  & 10,30  &  25.198  &  24.587  &   1.151  &	0.812  &   $-$  &  0.039  &  \\
V373 & 64629  & ab  & 0.606  &  137.3141  &  9,29  &  25.092  &  24.496  &   0.723  &	0.569  &   $-$  &  0.025  &  \\
V239 & 42860  & ab  & 0.608  &  136.7378  & 10,34  &  25.406  &  24.738  &   0.728  &	0.721  &   $-$  &  0.040  &  r\\
V244 & 43785  & ab  & 0.618  &  136.6791  & 11,33  &  24.988  &  24.306  &   0.956  &	0.777  &  0.054 &  0.038  &  r,o\\
V71  & 13845  & ab  & 0.619  &  136.9440  & 11,35  &  25.442  &  24.799  &   1.135  &	0.691  &  0.016 &  0.050  &  r\\	
V16  & 3518   & ab  & 0.660  &  136.5867  &  9,32  &  25.026: &  24.440: &$>$0.942  &$>$0.552  &   $-$  &  0.043  &  \\
V368 & 63851  & ab  & 0.661  &  137.3691  & 10,34  &  25.133  &  24.566  &   0.814  &	0.706  &   $-$  &  0.052  &  \\
V363 & 62059  & ab  & 0.669  &  136.7334  & 11,33  &  24.924: &  24.541: &$>$0.854  &$>$0.690  &  0.046 &  0.058  &  b\\
V252 & lc4324 & ab  & 0.522  &  138.5254  & 10,35  &  25.438  &  24.693  &   1.247  &	0.901  &   $-$  &  0.061  & u \\
V20  & lc1109 & ab  & 0.542  &  142.6077  & 10,35  &  25.462  &  24.687  &   0.900  &	0.871  &   $-$  &  0.039  & u \\
V273 & lc1418 & ab  & 0.579  &  138.5501  & 10,33  &  25.436  &  24.637  &   1.261  &	0.762  &   $-$  &  0.040  & u \\
V291 & lc2607 & ab  & 0.597  &  138.6625  & 11,27  &  25.345  &  24.857  &   1.000  &	0.854  &   $-$  &  0.034  & u \\
V236 & lc1434 & ab  & 0.600  &  136.8293  & 11,34  &  25.338  &  24.891  &   0.887  &	0.702  &   $-$  &  0.043  & u \\
V173 & lc453  & ab  & 0.603  &  137.6550  & 10,35  &  25.063  &  24.517  &   0.905  &	0.601  &   $-$  &  0.032  & u \\
V118 & lc3704 & ab  & 0.611  &  138.5208  & 11,33  &  25.492  &  24.809  &   1.039  &	0.667  &  0.038 &  0.038  & u \\
V251 & lc1611 & ab  & 0.680  &  136.7060  & 10,35  &  24.956: &  24.581: &$>$0.601  &$>$0.584  &   $-$  &  0.042  & u \\
\hline
\end{tabular}
\end{center}

r = outside the red edge of the RR Lyrae instability strip\\
b = outside the blue edge of the RR Lyrae instability strip\\
o = overluminous\\
u = uncertain. These 8 {\it ab-}type RR Lyrae stars have been approximately 
scaled following the procedure described in Sect.~\ref{sez:fluxes}.

\end{table*} 

\begin{table*}
\begin{center}
\caption{Basic properties of Cepheids}
\label{tab:cc}													        
\begin{tabular}{lrrc cc@{\hspace{20pt}}l @{\hspace{10pt}}lrr ccc}
\hline
IDr & ID    & ~~~Type &   $P$    & Epoch     &  $N$   &$<B_{int}>$ &$<V_{int}>$&
$A_B$~~~&$A_V$~~~& $resB$  & $resV$ & Notes \\
 &     &      & (days) &($-$2452000)& $(B,V)$ & (mag)   & (mag)    &  (mag)  &  (mag)  & (mag) & (mag) & \\
\hline
V28  & 6555  & LL-C &  0.364  & 136.8996  & 10,30  & 24.730  &  23.968  &   0.568  &   0.256  &  $-$   & 0.029  &  r\\
V195 & 33491 & LL-C &  0.427  & 136.5877  & 10,32  & 23.960  &  23.214  &   0.256  &   0.168  &  $-$   & 0.012  &  \\
V287 & 49919 & ~~~C &  0.444  & 137.3166  & 11,34  & 23.011  &  22.414  &   0.293  &   0.326  & 0.021  & 0.013  &  \\
V352 & 60392 & LL-C &  0.485  & 137.1243  & 11,34  & 24.983  &  24.280  &   1.170  &   0.619  & 0.172  & 0.030  &  r\\
V289 & 50637 & ~~~C &  0.582  & 136.7562  & 10,30  & 23.324: &  22.390: &$>$0.356  &$>$0.184  &  $-$   & 0.007  &  r\\
V321 & 54941 & LL-C &  0.589  & 136.7459  & 11,35  & 24.831  &  24.028  &   0.718  &   0.473  & 0.055  & 0.026  &  r\\
V298 & 51755 & ~~~C &  0.593  & 136.3162  & 11,34  & 23.158  &  22.215  &   0.181  &   0.139  & 0.008  & 0.004  &  r\\
V36  & 7820  & LL-C &  0.594  & 136.5410  & 10,34  & 25.021  &  24.237  &   0.618  &	0.480  &  $-$	& 0.033  &  r\\
V343 & 59491 & LL-C &  0.594  & 136.0593  &  9,31  & 24.861  &  23.762  &   0.569  &   0.245  &  $-$	& 0.022  &  rr\\
V70  & 14051 & LL-C &  0.596  & 136.7500  & 11,34  & 25.039  &  24.239  &   0.665  &	0.325  & 0.042  & 0.017  &  r\\
V27  & 6138  & LL-C &  0.598  & 136.3832  & 11,33  & 24.153  &  23.219  &   0.900  &	0.495  & 0.074  & 0.043  &  rr\\
V63  & 12626 & LL-C &  0.599  & 136.5500  & 10,32  & 24.932  &  24.220  &   0.606  &	0.253  &  $-$	& 0.012  &  r\\
V382 & 66108 & LL-C &  0.603  & 136.7785  & 11,34  & 24.593  &  23.601  &   0.594  &	0.290  & 0.082  & 0.014  &  rr\\
V149 & 26386 & LL-C &  0.604  & 142.6521  & 11,34  & 24.702  &  24.234  &   0.890  &	0.415  & 0.040  & 0.030  &  \\
V202 & 34124 & LL-C &  0.606  & 136.5086  & 11,33  & 24.643  &  23.903  &   0.529  &	0.381  & 0.048  & 0.012  &  r\\
V265 & 46406 & LL-C &  0.607  & 136.6344  & 11,34  & 24.003  &  23.358  &   0.613  &   0.354  & 0.049  & 0.019  &  \\
V260 & 45910 & ~~~C &  0.607  & 137.2264  & 11,35  & 23.507  &  22.882  &   0.839  &   0.654  & 0.007  & 0.013  &  \\
V229 & 40366 & LL-C &  0.612  & 136.9319  & 10,29  & 24.232  &  23.520  &   0.322  &	0.138  &  $-$	& 0.006  &  \\
V79  & 15029 & LL-C &  0.621  & 136.9059  & 10,30  & 24.500  &  23.836  &   0.245  &	0.132  &  $-$	& 0.011  &  \\
V83  & 15435 & ~~~C &  0.625  & 135.4346  & 11,35  & 23.387: &  22.766: &$>$0.607  &$>$0.460  & 0.008  & 0.014  &  \\
V388 & 67385 & LL-C &  0.629  & 138.5180  & 11,30  & 24.899: &  24.282: &$>$0.726  &$>$0.604  & 0.048  & 0.046  &  \\
V108 & 19458 & ~~~C &  0.654  & 136.5122  & 11,34  & 23.550  &  22.947  &   0.418  &   0.271  & 0.023  & 0.011  &  \\
V269 & 46990 & LL-C &  0.660  & 136.7447  & 11,33  & 24.747  &  24.175  &   0.604  &	0.546  & 0.079  & 0.054  &  \\
V286 & 49740 & LL-C &  0.670  & 136.4926  & 10,34  & 24.565  &  24.021  &   0.627  &	0.273  &  $-$	& 0.024  &  \\
V4   & 1453  & ~~~C &  0.706  & 142.6949  & 11,34  & 23.559  &  22.732  &   0.461  &   0.357  & 0.034  & 0.013  &  r\\
V290 & 50545 & LL-C &  0.708  & 136.6677  & 11,33  & 24.167: &  23.517: &$>$0.181  &$>$0.169  & 0.006  & 0.012  &  \\
V181 & 31770 & ~~~C &  0.741  & 136.7826  & 11,34  & 23.516  &  22.853  &   0.476  &   0.371  & 0.008  & 0.011  &  \\
V58  & 11820 & LL-C &  0.768  & 136.8652  & 11,35  & 23.792  &  23.180  &   1.111  &	0.839  & 0.014  & 0.017  &  \\
V390 & 67313 & ~~~C &  0.946  & 136.8249  & 11,34  & 23.088  &  22.507  &   0.746  &	0.415  & 0.016  & 0.017  &  \\
V315 & 53467 & ~~~C &  0.953  & 135.9134  & 11,34  & 23.080  &  22.404  &   0.774  &	0.601  & 0.009  & 0.015  &  \\
V278 & 48991 & ~~~C &  1.075  & 136.5400  & 11,34  & 22.804  &  22.114  &   0.311  &	0.309  & 0.007  & 0.009  &  \\
V358 & 61412 & ~~~C &  1.152  & 136.1990  & 11,32  & 23.114: &  22.484: &$>$0.733  &$>$0.618  & 0.009  & 0.022  &  \\
V364 & 62140 & ~~~C &  1.232  & 136.9550  & 11,35  & 22.644: &  22.068: &$>$0.387  &$>$0.295  & 0.011  & 0.012  &  \\
V369 & 64244 & ~~~C &  1.401  & 138.5944  & 11,34  & 23.323  &  22.538  &   1.583  &	0.984  & 0.061  & 0.035  &  \\
V183 & 31835 & ~~~C &  1.404  & 137.4288  & 11,35  & 22.485  &  21.895  &   0.719  &	0.492  & 0.007  & 0.007  &  \\
V386 & 66653 & ~~~C &  1.413  & 127.6621  & 11,33  & 23.394  &  22.649  &   0.955  &	0.672  & 0.032  & 0.023  &  \\
V276 & 48184 & ~~~C &  1.606  & 138.3337  & 11,35  & 23.074: &  22.409: &$>$1.250  &$>$0.943  & 0.011  & 0.011  &  \\
V105 & 18672 & ~~~C &  1.653  & 136.6989  & 11,32  & 22.762: &  22.184: &$>$0.383  &$>$0.363  & 0.006  & 0.004  &  \\
V140 & 25240 & ~~~C &  1.661  & 136.8237  & 11,35  & 23.121: &  22.516: &$>$1.326  &$>$0.856  & 0.011  & 0.009  &  \\  
V29  & 6869  & ~~~C &  1.674  & 136.5700  & 11,35  & 23.608: &  22.743: &$>$0.545  &$>$0.375  & 0.014  & 0.017  &  r\\
V255 & 45136 & ~~~C &  1.695  & 134.7770  & 11,35  & 23.220: &  22.475: &$>$0.846  &$>$0.525  & 0.025  & 0.008  &  \\ 
V340 & 58803 & ~~~C &  1.696  & 136.5150  & 11,35  & 23.198: &  22.442: &$>$0.824  &$>$0.626  & 0.023  & 0.016  &  \\
V380 & 66001 & ~~~C &  1.868  & 142.6077  & 11,33  & 23.138  &  22.353  &   0.517  &	0.555  & 0.040  & 0.015  &  \\
V40  & 8364  & ~~~C &  1.952  & 136.2943  & 11,32  & 23.184: &  22.426: &$>$0.821  &$>$0.643  & 0.005  & 0.008  &  \\
V370 & 64036 & ~~~C &  2.006  & 136.2700  & 11,31  & 22.881: &  22.197: &$>$0.699  &$>$0.838  & 0.010  & 0.017  &  \\
\hline
\end{tabular}
\end{center}

r =  outside the red edge of the instability strip\\
rr=  too red colors (stars in crowded regions close to the center 
of \mygal\ or near to regions of active star formation) 
\end{table*} 
\begin{table*}
\begin{center}
\caption{Basic information on eclipsing binaries}
\label{tab:bb}
\begin{tabular}{lrrc cc@{\hspace{20pt}}l @{\hspace{10pt}}lrr ccc}
\hline
IDr & ID    & ~~~Type &   $P$    & Epoch     &  $N$   &$<B_{int}>$ &$<V_{int}>$&
$A_B$~~~&$A_V$~~~& $resB$  & $resV$ & Notes \\
 &     &      & (days) &($-$2452000)& $(B,V)$ & (mag)   & (mag)    &  (mag)  &  (mag)  & (mag) & (mag) & \\
\hline
V205 & 34739 & EB  & 0.173  & 137.6105  &  11,35  &  23.179  &  23.117  & 0.432  &  0.326  & 0.016 &  0.022  & \\ 
V115 & 21503 & EB  & 0.221  & 138.5170  &  11,35  &  23.743  &  23.153  & 0.370  &  0.174  & 0.028 &  0.016  & \\
V24  & 5659  & EB  & 0.431  & 142.6410  &  11,33  &  21.329  &  21.225  & 0.402  &  0.402  & 0.007 &  0.008  & \\   
V274 & 48260 & EB  & 0.432  & 142.5515  &  11,35  &  23.116  &  22.911  & 0.469  &  0.237  & 0.032 &  0.017  & \\
V216 & 36305 & EB  & 0.588  & 142.4700  &   8,35  &  21.050  &  21.135  & 0.138  &  0.591  &  $-$  &  0.009  & \\
V121 & 22565 & EB  & 0.764  & 138.5501  &  10,33  &  25.318  &  24.742  & 2.134  &  2.539  &  $-$  &  0.116  & \\
\hline
\end{tabular}
\end{center}
\end{table*}

\begin{acknowledgements}
We warmly thank Marcella Marconi for providing in advance of
publication the theoretical boundaries for the instability strip of
RR Lyrae stars, Anomalous and Classical Cepheids calculated from the
latest version of her pulsation models. We thank the anonymous referee
for useful comments. 

\end{acknowledgements}

\appendix
                               
\section{Atlas of light curves} 
                  
 We present here an atlas of light curves for the  
variables stars in \mygal\ with light curves on a magnitude scale and
good sampling of the light variation.
The photometric data 
are folded according to the ephemerides given 
in Tables~\ref{tab:rr}, \ref{tab:cc}, \ref{tab:bb}.
Variable stars are grouped by type: RR Lyrae stars ({\it ab-, c-}type 
separately), Cepheids, and eclipsing
binaries. Within each 
group the variables are ordered by increasing period. 


\begin{figure*} 
\includegraphics[width=17cm, bb=45 144 457 718, draft=true]{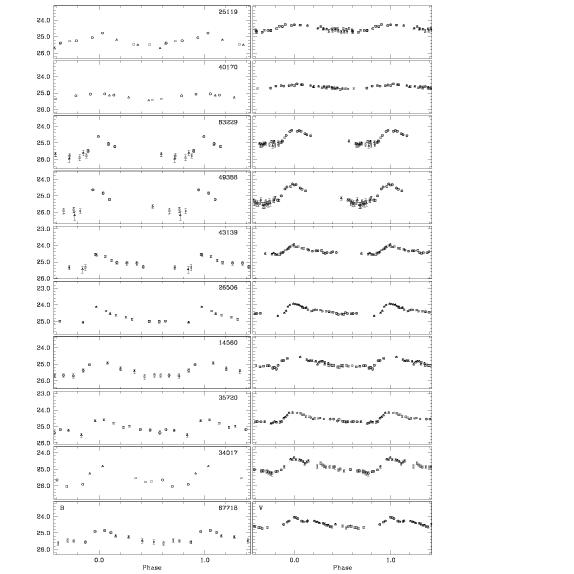}
\caption{$B,V$ light curves of the RR Lyrae stars in our field; the 
variable stars are ordered by increasing period.}
\label{fig:appb1}
\end{figure*}

\begin{figure*} 
\includegraphics[width=17cm, bb=45 252 457 718, draft=true]{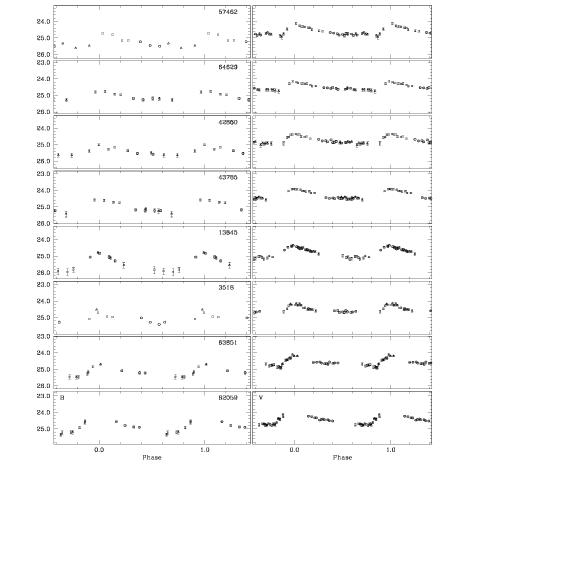}
{{\bf Fig.~\ref{fig:appb1}} -- continued --}
\end{figure*}

\begin{figure*} 
\includegraphics[width=17cm, bb=45 252 457 718, draft=true]{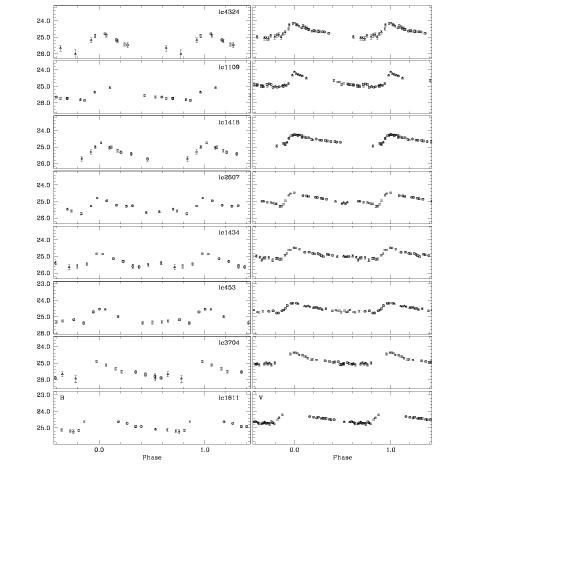}
\caption{$B,V$ light curves of the 8 likely $ab-$type RR Lyrae stars (see
Sect.~\ref{sez:fluxes}), ordered by increasing period.}
\label{fig:appb2}
\end{figure*}

\begin{figure*} 
\includegraphics[width=17cm, bb=45 144 457 718, draft=true]{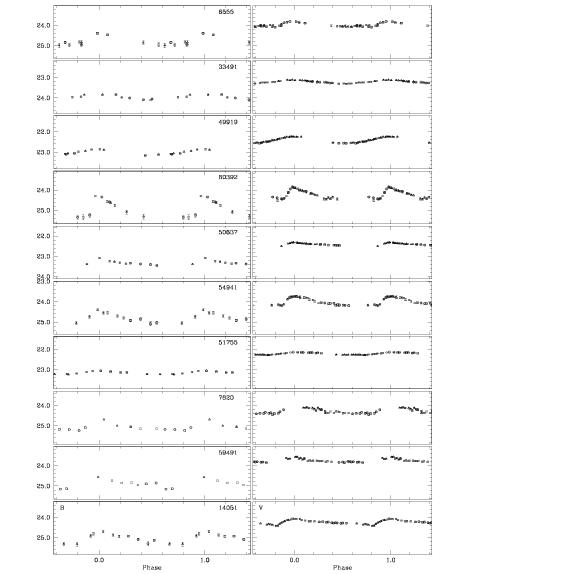}
\caption{$B,V$ light curves of Low Luminosity (LL) and Classical
Cepheids, ordered by increasing period.}
\label{fig:appb3}
\end{figure*}

\begin{figure*}
\includegraphics[width=17cm, bb=45 144 457 718, draft=true]{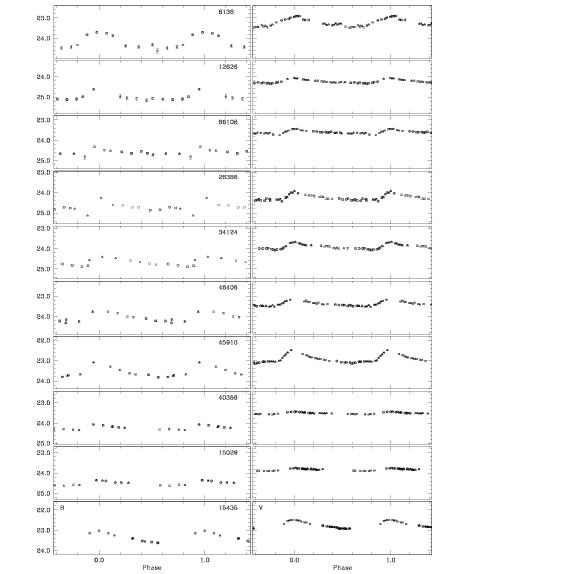}
{{\bf Fig.~\ref{fig:appb3}} -- continued --}
\end{figure*}

\begin{figure*} 
\includegraphics[width=17cm, bb=45 144 457 718, draft=true]{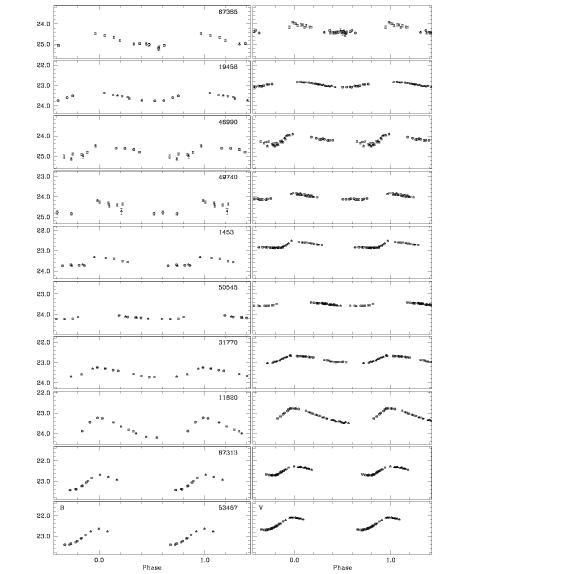}
{{\bf Fig.~\ref{fig:appb3}} -- continued --}
\end{figure*}

\begin{figure*} 
\includegraphics[width=17cm, bb=45 144 457 718, draft=true]{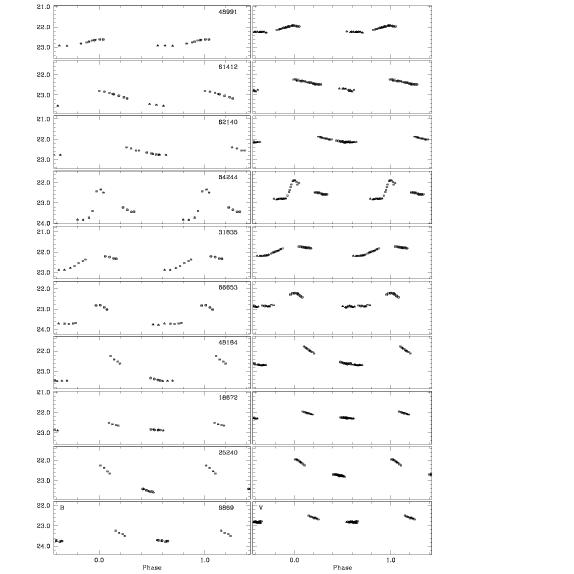}
{{\bf Fig.~\ref{fig:appb3}} -- continued --}
\end{figure*}

\begin{figure*} 
\includegraphics[width=17cm, bb=45 416 457 718, draft=true]{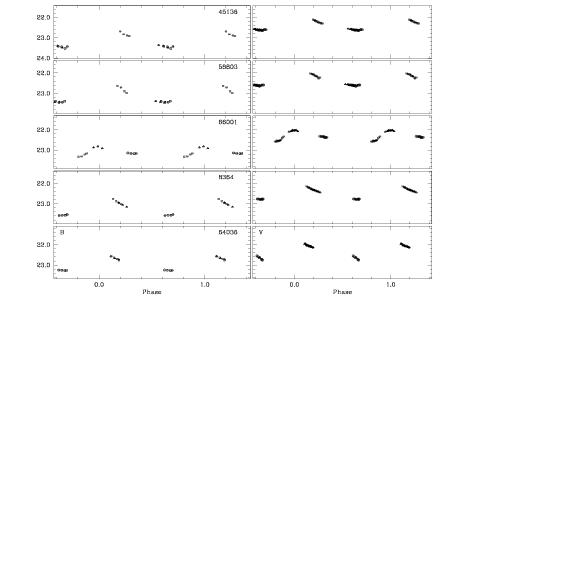}
{{\bf Fig.~\ref{fig:appb3}.} -- continued --}
\end{figure*}

\begin{figure*} 
\includegraphics[width=17cm, bb=45 364 457 718, draft=true]{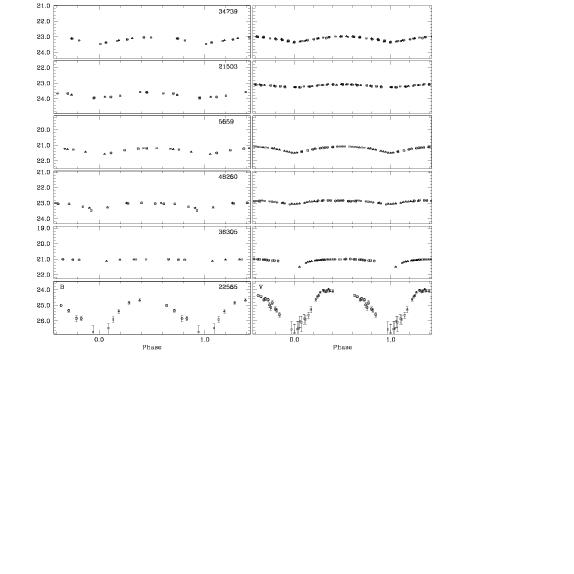}
\caption{$B,V$ light curves of eclipsing binaries, ordered 
by increasing period.}
\label{fig:appb4}
\end{figure*}

\end{document}